\documentclass[twocolumn,showpacs,preprintnumbers,amsmath,amssymb]{revtex4}
\date{\today}

\usepackage{graphicx}

\newcommand{\tE}{\tilde{E}}
\newcommand{\trho}{\tilde{\rho}}
\newcommand{\tkappa}{\tilde{\kappa}}
\newcommand{\tgamma}{\tilde{\gamma}}
\def\Re{{\bf Re}}
\begin{document}

\title{
A ring laser gyroscope without lock-in phenomenon
}

\author{Satoshi Sunada, Shuichi Tamura, Keizo Inagaki, and Takahisa Harayama}
\affiliation{
Department of Nonlinear Science ATR Wave Engineering Laboratories
2-2-2 Hikaridai Seika-cho Soraku-gun Kyoto 619-0228 Japan
}
\email{sunada@atr.jp}

\begin{abstract}
We theoretically and numerically study the effect of backscattering on 
rotating ring lasers
by employing the Maxwell-Bloch equations.
We show that frequency shifts due to the Sagnac effect incorporating the
 effect of backscattering can be observed without lock-in phenomenon, 
if the strength of backscattering originating
in the bumps of the refractive index is larger than a certain value.    
It is also shown that 
the experimental results corresponding to the theoretical ones 
can actually be obtained by using a semiconductor fiber-optic ring laser
 gyroscope.
\end{abstract}


\maketitle

\section{Introduction}
Ring laser gyroscopes (RLGs) operate laser action on 
counter-propagating waves 
and measure their frequency difference (beat frequency) due to 
the Sagnac effect \cite{Post,Crow,Aronowitz}. 
The main problems for operation are caused by
the effects of backscattering and nonlinearity inherent in a gain
medium. 
For instance, 
when the rotation rate of the RLG is slow, 
frequency-locking effect induced by backscattering occurs between
the counter-propagating waves and eliminates the beat frequency due
to the Sagnac effect (Sagnac beat frequency), 
which is well-known as the
lock-in phenomenon 
\cite{Crow,Aronowitz,Laserphysics1,Laserphysics2,Lamb2}.
In conventional He-Ne RLGs,
avoiding the lock-in phenomenon has been one of the important issues in
order to measure the beat frequency at slow rotation rates, and   
many investigations have been intensively done for this purpose. 

While conventional RLGs utilize a gaseous medium (He-Ne) as a
gain medium, 
a new type of RLG using solid
states and a semiconductor material  
have recently received research attention
\cite{B-YKim1,B-YKim2,B-YKim3,Klochan,Schwartz1,Schwartz2,Sorel,Cao,Inagaki,Harayama}.
As is well-known, 
the nonlinearity
 inherent in these gain media
leads to a spatial hole burning effect and 
suppresses one of the counter-propagating waves 
due to gain competition (in an ideal ring laser without backscattering
and noise sources), 
unlike in the case of gaseous medium \cite{Laserphysics1,Laserphysics2}.
While several theoretical and experimental works exist for
solidstate RLGs, many of the investigations have so far focused on overcoming
or controlling such a nonlinearity \cite{B-YKim1,Schwartz1,Schwartz2,Cao}. 
Meanwhile, backscatter-induced phenomenon, such
as lock-in, and the properties of the RLGs at slow rotation rates 
have not yet been examined in detail.
  
In this paper, we theoretically and numerically study
the effect of backscattering on the RLGs. 
In the description, we use the Maxwell-Bloch equations, which consists
of the Maxwell equations rewritten in 
a rotating frame of reference 
and the Bloch equations that
have been used as a simple model of solid-state gain media or as a semiconductor
material.   
We show that
 Sagnac beat frequencies can be observed without 
the lock-in phenomenon,
if the strength of backscattering originating in the bumps of the
refractive index is larger than a certain value.
In this case, we found that the rotation rate dependence of the 
beat frequency cannot be described by a
conventional theory of the Sagnac
effect, which gives a linear relation between the beat frequency and
the cavity rotation rate \cite{Post,Crow}. 
Instead, we demonstrate that 
it follows a novel formula of the Sagnac effect incorporating the
effect of backscattering, 
which have been derived in Ref. \cite{Lamb2,Spreeuw}.  
In addition, we report that 
such rotation rate dependence 
can actually be observed without the lock-in phenomenon 
in a semiconductor fiber-optic ring laser gyroscope
(S-FOG) \cite{Inagaki,Harayama}, 
which is composed of a semiconductor 
optical amplifier as a gain medium and optical fibers to form a ring cavity. 

This paper is organized as follows: 
In Sec. \ref{sec_theory}, we start with the Maxwell equations in
rotating frames of reference and review the 
eigenmodes of rotating ring cavities with backscattering 
to explain the Sagnac effect incorporating the effect of backscattering.
Then, the Maxwell equations coupled with the Bloch equations
(Maxwell-Bloch equations) 
are introduced to describe the nonlinear
interaction between the light field and a gain medium. 
We analyze the lock-in phenomenon by expanding 
the Maxwell-Bloch equations by the eigenmodes of a 
rotating ring cavity with backscattering and 
derive the critical backscattering strength, above which the lock-in
phenomenon does not occur.    
In Sec. \ref{sec_numerical}, numerical simulations confirm 
whether the Sagnac beat frequency is observed without 
lock-in phenomenon.  
The experimental results of S-FOG are reported in Sec. \ref{sec_exp}.  
Finally, the summary of this paper is provided in Sec. \ref{sec_sum}

\section{Theory \label{sec_theory}}
 
Applying the electromagnetic equation of a naturally covariant form 
(or the general theory of relativity) 
to rotating ring cavities 
yields the Maxwell equations generalized to a non-inertial frame of 
reference in uniform rotation with angular velocity vector \cite{Lamb2,Anderson,Landau,Heer}. 
Assuming that the vector is perpendicular to the plane where there is
a ring cavity and that the light propagates one-dimensionally along the ring
cavity, the following equation can be obtained: 
\begin{eqnarray}
& &
\left(
\dfrac{\partial^2}{\partial s^2}
-
\dfrac{n^2(s)}{c^2}\dfrac{\partial^2}{\partial t^2}
\right)E(s,t)
- 
2\dfrac{R\Omega}{c^2}\dfrac{\partial^2}{\partial s\partial t}E(s,t)\nonumber \\
& & =2\beta(s)\dfrac{\partial}{\partial t}E(s,t)
+ 
\dfrac{4\pi n^2}{c^2}\dfrac{\partial^2}{\partial t^2}P(s,t) + F_1(s,t),
\label{Eeq}
\end{eqnarray}
where $s$ is the coordinate along the ring cavity in a rotating frame of
reference with angular velocity $\Omega$. 
$R(=L/(2\pi))$ denotes the radius of the ring cavity. 
$P(s,t)$ is the polarization depending on the nonlinear response of a
gain medium inside the cavity, whose expression is given in 
subsection \ref{sub_gain}.   
Noise $F_1$ is phenomenologically introduced 
in order to represent thermal
fluctuation, quantum noise, cavity vibrations and so on;
$
<F_1(s,t)>=0
$ and
$
<F_1(s,t)F_1^*(s',t')>=\sigma_1\delta(s-s')\delta(t-t'),
$ where $\sigma_1$ is the strength of the fluctuation, 
and $<\cdots>$ denotes a spatial-average or a time-average.
$c$ is the velocity of the light, and 
$n(s)$  and $\beta(s)$ respectively denote the refractive index and 
the background absorption. 
Throughout this paper, 
it is assumed that the effect of backscattering 
from one of rotating waves to the
opposite rotating waves can be described by 
the bumps of the refractive index and absorption.

\subsection{Eigenmodes of a rotating ring cavities with backscattering
\label{subsec:eigen}}
A key to
understanding the effects of backscattering and rotation 
is to analyze the eigenmode structure of the ring cavities 
\cite{Spreeuw,Neelen}.
The eigenmodes in a ring cavity can be described by the
following wave equation derived from the Maxwell Eq. (\ref{Eeq})
omitting absorption $\beta$, polarization term $P$, and noise
term $F_1$: 
\begin{eqnarray}
\left(
\dfrac{\partial^2}{\partial s^2} 
+ 
n^2(s) \dfrac{\omega^2(\Omega)}{c^2}
\right)U(s)
+
2i\dfrac{R\Omega\omega(\Omega)}{c^2}
\dfrac{\partial}{\partial s} U(s)
= 0, \label{fundeq}
\end{eqnarray}
where $E(s,t)$ oscillates with frequency $\omega(\Omega)$ as 
$E(s,t)=U(s)e^{-i\omega(\Omega) t}+c.c$ and 
$U$ is the wavefunction of the mode obtained
 under periodic boundary condition
$U(s+L)=U(s)$, where $L$ is the cavity length.
Below, it is assumed that 
the wavefunctions satisfy orthogonal relation
$
1/(n^2_0L)\oint U_i U_j^* n^2 ds = \delta_{ij} 
$ 
up to the first order of $R\Omega/c$.
In the above, $L$ is the total length of a ring cavity, 
and $n_0$ is spatial-averaged refractive index $n^2_0=1/L\oint n^2(s)ds$.

In the case of non-rotating ring cavity $\Omega=0$, 
Eq. (\ref{fundeq}) can
be reduced to a usual wave equation 
$
(\partial^2/\partial s^2 + n^2(s)\omega_j(0)^2/c^2)U_j^0(s)=0,
$
where $\omega_j(0)$ and $U_j^0$ respectively denote  
the eigen frequency and the wavefunction of a non-rotating cavity.
In particular, when the refractive index is uniformly distributed inside
cavity $n(s)=n_0$, 
 it is known that the solutions of the wave equation are 
clockwise (CW-) rotating wave function
 and counter-clockwise (CCW-) rotating wave
function with an identical frequency
$\omega_0=(ck_0=)2\pi mc/(n_0L)$, where $m$ is an integer ($m=1,2,..$).
 
%
However, such modes with identical
frequency cannot generally be
described as eigenmodes when the spatial distribution of refractive 
index $n(s)$ is non-uniform, i.e., there is a bump of the
refractive index in the ring cavity: $\Delta n^2(s)(=n^2-n_0^2)\ne 0$. 
In this case, the bump of the refractive index causes 
a linear coupling between the CW- and CCW- rotating wave modes due to
backscattering and the wavefunctions of the eigenmodes change into two standing
waves \cite{Spreeuw,Neelen,Etrich}. 
Then, the degenerate frequency is split into two frequencies.
For instance, when
the bumps of refractive index 
$\Delta n^2\ll 1$ are so small, the frequency splitting 
$\Delta \omega_0[=\omega_1(0)-\omega_2(0)]$ 
can be associated with the
strength of the backscattering which has a conservative nature,
as follows \cite{Spreeuw,Neelen}: 
\begin{eqnarray}
\Delta f_0 =\dfrac{\Delta\omega_0}{2\pi}\approx 
\dfrac{c}{n_0\pi L}\sqrt{\gamma} \hspace{1cm} (Hz), \label{f0} 
\end{eqnarray}
where $\gamma$ is the
intensity-reflection coefficient of the backscatters ($\gamma\ll 1$).
Then the two standing wave modes with the splitting eigen frequencies 
are described as follows;
\begin{eqnarray}
U_1^0(s)=\sqrt{2}\cos n_0 k_0 s, U_2^0(s)=\sqrt{2}\sin n_0 k_0 s. \label{pur_wf}
\end{eqnarray} 

Moreover, when the cavity is rotated, 
the rotation affects the standing wave modes of a non-rotating cavity and 
causes shifts of their eigen frequencies. 
The effects 
can be shown by applying the perturbation
theory for nearly-degenerate states typically used in quantum mechanics 
to Eq. (\ref{fundeq}) \cite{SH1,SH2,SH3}. 
Based on the perturbation theory, 
the eigenmodes in a rotating cavity with backscattering
can be represented as the superposition of the nearly-degenerate modes
as 
 \begin{eqnarray}
U = c_1U_1^0(s) + c_2U_2^0(s), \label{super}
\end{eqnarray}
where the ratio of coefficients $c_1$ and $c_2$ is obtained as
\begin{eqnarray}
c_2/c_1  = i\dfrac{\hat{S}\Omega}{\pm \sqrt{\hat{S}^2\Omega^2+\Delta\omega_0^2}
-\Delta\omega_0}. \label{ratio}
\end{eqnarray}
Then frequency $\omega_{j}(\Omega)$  
of eigenmode $j$ ($j=1,2$)
newly produced by cavity rotation up to the first order of $R\Omega/c$
is described as 
\begin{eqnarray}
\omega_{1}(\Omega) = \omega_0 + \dfrac{1}{2}
\sqrt{
\hat{S}^2\Omega^2 
+
\Delta \omega^2_0
}, \label{eq_waven1} \\
\omega_{2}(\Omega) = \omega_0 - \dfrac{1}{2}
\sqrt{
\hat{S}^2\Omega^2 
+
\Delta \omega^2_0
}, \label{eq_waven2}
\end{eqnarray}
where $\Delta \omega_0$ is the frequency difference between modes of
the non-rotating cavity and 
$\hat{S}$ is so-called scale factor, which can be 
represented as 
$\hat{S}=\frac{1}{n^2_0c\pi}
|\oint U_1^0\frac{\partial}{\partial s}U_2^0 ds|\approx
2\omega_0R/(n_0c)=4\omega_0A/(n_0cL) $ \cite{Crow}.  
From Eqs. (\ref{eq_waven1}) and (\ref{eq_waven2}), we can finally obtain 
frequency difference $\Delta f_{\Omega} = 
|\Delta\omega_2(\Omega)-\Delta\omega_1(\Omega)|/(2\pi)$ 
as follows:
\begin{eqnarray}
\Delta f_{\Omega} = \dfrac{\Delta \omega_{\Omega}}{2\pi} 
= \sqrt{S^2\Omega^2 + \Delta f^2_0
}\hspace{1cm} (Hz), \label{freqd}
\end{eqnarray}
where 
$S = \hat{S}/(2\pi)$ and $\Delta
f_0 = \Delta \omega_0/(2\pi)$.

In conventional theory of the Sagnac effect, 
frequency difference proportional to the rotation rate has been derived 
between the CW- and CCW- rotating wave modes 
by assuming that there are no backscatters in a ring cavity
\cite{Post,Crow}.
However, in the case of a ring cavity with backscattering, 
as seen in Eq. (\ref{freqd}), 
the relation between frequency difference $\Delta f_{\Omega}$ 
and rotation rate $\Omega$ is not proportional, and 
it is modified by the effect of backscattering $\gamma$.
Then as seen in Eq. (\ref{super}), the modes are
expressed as superposed standing waves with the ratio shown in Eq. (\ref{ratio}).
This is the Sagnac effect in a ring cavity with backscattering. 

\subsection{Model of a gain medium \label{sub_gain}}
Next, we discuss the case of a ring laser taking into account 
the effect of a gain medium in order to examine how the Sagnac effect is
observed in this case. 

To describe a gain medium in interaction with the light
field, we use the well-known optical Bloch equations, which
have been employed as a simple model of solid-state gain medium or 
as semiconductor gain medium:   
\begin{eqnarray}
\dfrac{\partial}{\partial t}\rho(s,t) 
&=& 
-(
\gamma_{\perp} + i\omega_t
)\rho(s,t) \nonumber \\
&-&i\kappa W(s,t)E(s,t) + F_2(s,t), \label{Peq}
\end{eqnarray}
\begin{eqnarray}
& &\dfrac{\partial}{\partial t}W(s,t) 
= 
-\gamma_{//}\left(
W(s,t)-W_{\infty}
\right) \nonumber \\ 
&-&2i\kappa E(s,t) 
\left(
\rho(s,t)-\rho^*(s,t)
\right) + F_3(s,t), \label{Weq}
\end{eqnarray}
where it is assumed that the medium is not influenced by
rotation. 
In the above, $\rho$ and $W$ denote microscopic polarization and population
inversion, respectively.
The two relaxation parameters, $\gamma_{\perp}$ and $\gamma_{//}$, 
 are the transversal relaxation and 
longitudinal relaxation rates, respectively. 
$\omega_t$ is the transition frequency between two level medium, and
$W_{\infty}$ is the pumping power.
Noises $F_2$ and $F_3$ are introduced phenomenologically 
in order to represent thermal and pumping fluctuations:
 \begin{eqnarray}
<F_i(s,t)> &=&0, \nonumber \\
<F_i(s,t)F_j^*(s',t')> &=& 
\sigma_i\delta_{ij}\delta(s-s')\delta(t-t'),
\end{eqnarray}
where $\sigma_i$ ($i=1,2,3$) 
is fluctuation strength.
%
%
The Bloch equations (\ref{Peq}) and (\ref{Weq}) 
are coupled with the Maxwell Eq. (\ref{Eeq}) through the following relation
between microscopic polarization $\rho$ and macroscopic one $P$:
%
\begin{equation}
 P(s,t) = N(s) \left( \rho(s,t) + \rho^*(s,t) \right) \kappa \hbar, 
\label{MB-Pz}
\end{equation}
where $N(s)$ is the atomic number density described as 
$N(s)=N_a\Theta(s)$,
where $\Theta(s)$ is a step function that is $1$
inside a region of a gain medium and zero outside
it. 

Equations (\ref{Eeq}), (\ref{Peq}), (\ref{Weq}), and (\ref{MB-Pz}) are 
the fundamental ones describing the light field and the gain medium
mentioned above. 
\subsection{Two mode operation in rotating ring lasers
  \label{subsec_theo}}
%
%
In analyzing the dynamics of 
Eqs. (\ref{Eeq}), (\ref{Peq}), (\ref{Weq}), and (\ref{MB-Pz}) 
in ring lasers, we expand these equations by the eigenmode
basis of a rotating cavity derived in subsection \ref{subsec:eigen} and 
derive the amplitude- and the phase- equations of the modes,
although in conventional theoretical approaches, CW- and CCW- rotating
wave basis has been frequently used \cite{Crow,Aronowitz,Laserphysics1,Laserphysics2,Lamb2,Klochan,Schwartz1,Schwartz2,Sorel}. 
As demonstrated in Ref. \cite{Spreeuw,Neelen,Etrich}, 
using the eigenmode basis makes it possible to 
simply discuss dynamical behaviors of ring lasers.

To focus on the effect of a gain medium, 
we here omit noise terms $F_i$ ($i=1,2,3$) in the Maxwell-Bloch
equations (\ref{Eeq}), (\ref{Peq}), (\ref{Weq}), and (\ref{MB-Pz}), and assume that electric
field $E(s,t)$ are described by two eigenmodes $U_j$ ($j=1,2$) as
follows:
\begin{eqnarray}
E(s,t)&=&\tE(s,t) e^{-i\omega_0 t} +c.c  \label{Eepx} \\
&=& \sum_{j=1,2}E_j(t)e^{i\phi_j(t)}U_j(s)e^{-i\omega_0t} +c.c\nonumber,
\end{eqnarray}
where 
$E_j$ and $\phi_j$ $\in \Re$ denote the amplitude and the phase of the mode
$j$ of a rotating ring cavity, respectively. 
Then we assume that frequency 
$\omega_0$ is close to the
transition frequency $\omega_t$ of the gain medium,
$|\omega_0-\omega_t|/\omega_t\ll 1$. 

In this case, 
solving Eqs. (\ref{Peq}), (\ref{Weq}), and (\ref{MB-Pz}) by a perturbational method 
for small electric field yields 
$P \approx -i\alpha_0\tE W\Theta + c.c$ and 
$
W = W_{\infty}
(1-\beta_0|\tE|^2) +
O(|\tE|^4),
$
where $\alpha_0=2\pi N_a\kappa^2\hbar\omega_0/\gamma_{\perp}$ and 
$\beta_0 = 4\kappa^2/(\gamma_{\perp}\gamma_{//})$.
By substituting them in the Maxwell equation (\ref{Eeq}) and using the
following orthogonal relation
$
1/(n^2_0L)\oint U_i^* U_j n^2(s)ds 
= 
\delta_{ij} 
+ 
O\left(
|R\Omega/c|^2
\right) 
$, 
where $n_0$ is the spatial-averaged refractive index, 
$n^2_0=1/L\oint n^2 ds$,
we can obtain the following equations for amplitude $E_j$;
\begin{eqnarray}
\dfrac{d}{dt}&E_j& = 
\left(
g_j - s_j E_j^2 -c_{12} E_{3-j}^2 
\right)E_j \nonumber \\
&-& M_j^c \cos\Psi -M_j^s \sin\Psi 
- \bar{M}_j \cos(2\Psi +\theta_{\chi}),
\label{mode-eq1}
\end{eqnarray}
where ``net-gain'' coefficient $g_j$, self-saturation
coefficient $s_j$, and cross-saturation coefficient $c_{12}$ 
are given in Table \ref{tab:Table1}. 
$\Psi$ is the phase difference, i.e., $\phi_2-\phi_1$.
Modulation amplitudes $M_j^c$, $M_j^s$, and $\bar{M}_j$ are described
as: 
\begin{eqnarray}
M_j^c =
(
|\gamma_{12}|\cos\theta_{\gamma}
+
3|\xi_j|\cos\theta_j E_j^2 \nonumber \\
+
|\xi_{3-j}|\cos\theta_{3-j} E_{3-j}^2
)E_{3-j}, \label{modulation1}
\end{eqnarray}
\begin{eqnarray}
M_j^s =
(-1)^{3-j}
(
|\gamma_{12}|\sin \theta_{\gamma}
+
3|\xi_j|\sin\theta_j E_j^2 \nonumber \\
-
|\xi_{3-j}|\sin\theta_{3-j}E_{3-j}^2
)E_{3-j}, \label{modulation2}
\end{eqnarray}
\begin{eqnarray}
\bar{M}_j =
|\chi |E_{3-j}^2E_{j},
\end{eqnarray}
where 
$\theta_{\gamma}$, $\theta_j$, and $\theta_{\chi}$
are the arguments of $\gamma_{12}$, $\xi_j$, and
$\chi$, respectively.

On the other hand, 
the dynamics of phase difference $\Psi(=\phi_2-\phi_1)$ is 
described as:
\begin{eqnarray}
\dfrac{d}{dt}\Psi = -\Delta\omega_{\Omega} + H(\Psi), \label{mode-eq2}
\end{eqnarray}
where 
$\Delta\omega_{\Omega}(=\omega_2(\Omega)-\omega_1(\Omega))$ 
is the frequency difference between the two modes, as shown in
Eq. (\ref{freqd}). 
$H$, which is a periodic function of $\Psi$ (period $2\pi$), 
can be described as follows:
\begin{eqnarray}
H(\Psi)
&=&
|\gamma_{12}|\dfrac{E_1^2+E_2^2}{E_1E_2}
\sin(\Psi+\theta_{\gamma})  \nonumber \\
&+&
|\chi |
(E_1^2+E_2^2)
\sin(2\Psi+\theta_{\chi}),
\label{Heq} 
\end{eqnarray}
when amplitudes $E_1$ and $E_2$ are small enough and
$\xi_1\approx \xi^*_2 \ll 1$.

The solutions of Eqs. (\ref{mode-eq1}) and (\ref{mode-eq2}) can be
mainly classified into two: one is a stable stationary solution with 
constant amplitudes and
a constant phase difference, and the other is with
 time-dependent phase difference and amplitude modulating with
the phase difference. 
The distinction can be clearly made by examining Eq. (\ref{mode-eq2}),
which describes the time evolution of the phase difference. 

Note that the form of Eq. (\ref{mode-eq2}) resembles
with the well-known locking equation
\cite{Crow,Aronowitz,Laserphysics1,Laserphysics2,Lamb2}. 
This means that 
the solution of Eq. (\ref{mode-eq2}) markedly depends 
on the relative strengths of $\Delta\omega_{\Omega}$ and
$\max{|H(\Psi)|}$, 
where $\max{|H(\Psi)|}$ denotes the maximal value of $|H(\Psi)|$, 
which is obtained by solving the stationary equations of $dE_i/dt =0$.    
For instance, 
in the case of $\Delta\omega_{\Omega} < \max{|H(\Psi)|}$,  
Eq. (\ref{mode-eq2}) has stable fix points given by the solution of
$d{\Psi}/dt=0$. 
According to definition of $\Psi$, condition $d{\Psi}/dt=0$ implies that the
frequency difference between modes $1$ and $2$ equals zero, 
i.e., the lasing frequency of mode $1$ becomes 
identical with that of mode $2$, which is the lock-in 
(frequency locking) phenomenon.
%
On the other hand, when condition $\Delta\omega_{\Omega} \gg
\max{|H(\Psi)|}$ is satisfied, 
the lock-in phenomenon does not occur.  
Thus, phase difference $\Psi$ can be approximately described as 
$\Psi(t)\approx-\Delta\omega_{\Omega} t + \Psi_0$, 
where $\Psi_0$ is an initial phase, while 
the amplitudes $E_j$ $(j=1,2)$ of the lasing modes oscillate 
with frequency $\Delta\omega_{\Omega}$ 
through the modulation terms of Eq. (\ref{mode-eq1}) \cite{Harayama}.


Here we emphasize that frequency difference
$\Delta\omega_{\Omega}$ in Eq. (\ref{mode-eq2}) not only depends on 
rotation rate $\Omega$ but also on the strength of conservative
backscattering $\gamma$. As shown in Eq. (\ref{freqd}),
$\Delta\omega_{\Omega}$ never becomes zero even at rotation rate $\Omega=0$
when there are backscatters inside the ring cavity.
This means that the lock-in phenomenon does not occur even at 
rotation rate $\Omega=0$
if frequency splitting $\Delta\omega_{0}$ is large enough to
satisfy condition:
\begin{eqnarray}
\Delta\omega_0
>
\max{|H(\Psi)|_{\Omega=0}} \equiv \Delta\omega_{c}. \label{eq:condition} 
\end{eqnarray}
This condition is rewritten using Eq. (\ref{f0}) as follow:
 \begin{eqnarray}
\gamma > \gamma_c = \dfrac{n^2_0L^2}{4c^2}\Delta\omega^2_{c}. 
\label{eq:condition2} 
\end{eqnarray}
Although criterion $\Delta\omega_c$ is too complicated to be analytically
expressed for general cases, 
it has the simple form for a case where 
the wavefunctions of the two modes can be expressed by
Eq. (\ref{pur_wf}) because of $\gamma \ll 1$ and where 
 wavelength $2\pi/k_0$ of the light is shorter than the size of the ring
 cavity $k_0\int \Theta(s) ds \gg 1$. 
Therefore $g_1\approx g_2 = g$ and $s_1\approx s_2 =s$.
By solving equation $dE_i/dt=0$ under such conditions, 
 we can derive criterion $\Delta\omega_c$:
\begin{eqnarray}
\Delta\omega_c = 2\max
\left|
\dfrac{|\gamma_{12}|(K-1)\sin\Psi + g\sin 2\Psi}
{K+\cos 2\Psi}
\right|, \label{eqh}
\end{eqnarray}
where 
$K= s/\chi + 2\approx 5$.
In the case of (dispersive coupling) $\gamma_{12}$ 
$\gg g$ (linear net gain), since $H(\Psi)$ has maximal value at
$\Psi_m=\pi/2(2m+1)$, where $m$ is an integer, Eq. (\ref{eqh}) is reduced
to $\Delta\omega_c = 2|\gamma_{12}|$, while in the case of $\gamma_{12}
\ll g$, 
it is reduced to  
$\Delta\omega_c = 2g/\sqrt{K^2-1}$ \cite{Neelen_ex}, where
$H(\Psi)$ takes the maximal value at $\Psi_m = m\pi
-1/2\arctan\sqrt{K^2-1}$. 

So far, it has been believed that 
the lock-in phenomenon inevitably occurs at slow rotation
rates in a ring laser with backscattering. 
However, this result means that,
if conditions
(\ref{eq:condition}) or (\ref{eq:condition2}) are satisfied,  
the ring laser 
will have a property that 
the lock-in phenomenon can be naturally avoided without
 special techniques used 
in conventional RLGs, such as dithering \cite{Crow}.  
It can be concluded that 
the frequency difference $\Delta\omega_{\Omega}$ 
described by Eq. (\ref{freqd}) without the lock-in phenomenon can
be observed even at arbitrary slow rotation rate.

\begin{table}[h]
{\bf \caption{Coefficients in Eqs. (\ref{mode-eq1}), (\ref{modulation1}),
 (\ref{modulation2}), and (\ref{mode-eq2}).}
\label{tab:Table1}
}\begin{center}
\begin{tabular}{|l|l|}\hline
Coefficient & Physical Content \\ \hline
\rule[-1ex]{0pt}{3.5ex}
$g_j = \alpha_0 t_j W_{\infty} - \gamma_{jj}$ 
&
 Linear net gain \\
\rule[-1ex]{0pt}{3.5ex}
$s_i=\dfrac{\beta_0 W_{\infty}}{n_0^2 L}\oint\Theta |U_i|^4 n^2ds$ 
& 
Self-saturation\\
%
%
\rule[-1ex]{0pt}{3.5ex}
$c_{12}=\dfrac{2\beta_0W_{\infty}}{n_0^2L}\oint \Theta |U_1|^2|U_2|^2n^2 ds$
& 
Cross-saturation\\
%
%
\rule[-1ex]{0pt}{3.5ex}
$\xi_j=\dfrac{\beta_0W_{\infty}}{n_0^2L}\oint \Theta |U_j|^2U_i^*U^2_{3-j}n^2 ds$
& 
Modulation\\
%
%
\rule[-1ex]{0pt}{3.5ex}
$\chi=\dfrac{\beta_0W_{\infty}}{n_0^2L}\oint \Theta U^{*2}_1U_2^2n^2 ds$
& 
Modulation\\
%
%
\rule[-1ex]{0pt}{3.5ex}  
$\alpha_0 = \dfrac{2\pi N_a\kappa^2\hbar \omega_0}{\gamma_{\perp}}$
&   
First-order factor
\\
\rule[-1ex]{0pt}{3.5ex}  
$t_j=\dfrac{1}{n_0^2 L}\oint \Theta |U_i|^2 n^2ds$ 
&
Light-medium coupling 
\\
%
%
\rule[-1ex]{0pt}{3.5ex}
$\gamma_{ij} = \dfrac{c^2}{n_0^2 L}\oint\beta U_i^* U_j ds$
 &
Dispersive coupling
\\
\rule[-1ex]{0pt}{3.5ex}  
$\beta_0=\dfrac{4\alpha_0 \kappa^2}{\gamma_{\perp}\gamma_{//}}$ &
Third-order factor\\
%
%
\rule[-1ex]{0pt}{3.5ex}  
$n^2_0=\dfrac{1}{L}\oint n^2 ds$ &
 Averaged refractive index \\
\hline
\end{tabular}
\end{center}
\end{table}

\section{Numerical simulations \label{sec_numerical}}%
%
In the previous section, we derived the amplitude and the
phase-difference equations under two mode operation 
by expanding the Maxwell-Bloch equations 
by the eigenmodes of a rotating ring cavity with backscattering, 
and discuss the occurrence of lock-in phenomenon.  
Note that a perturbation theory for $\rho$ and $W$ 
and several approximations, including $E_i\ll 1$, are employed in the
derivation. 
Additionally, noise terms $F_i$ are omitted.
Thus, it is needed to check the validity of the
theoretical predictions obtained in the previous subsection 
by the Maxwell-Bloch Eqs. (\ref{Eeq}), (\ref{Peq})-
(\ref{MB-Pz})
including noise terms and full-order nonlinearity of a gain medium. 
In this section, we restrict ourselves to discuss the lock-in phenomenon 
by the numerical simulations. 
Detailed discussions of the effect of noises on lasing dynamics 
will be reported elsewhere. 

Note that it will take a very long time to simulation the Maxwell-Bloch
equations because the fast oscillation as well as 
the slowly varying envelop of the light field are simulated. 
The fast oscillation part of the light field has the constant frequency,
which is very close to the transition frequency $\omega_t$.
Thus it is possible to separate the slowly varying part from the fast
oscillation part in order to investigate the dynamics of rotating ring
lasers. 

Let us suppose $\tE$ and $\trho$ to be the slowly varying envelope 
of the electric field and the polarization field as 
$E=\tE e^{-i\omega_0 t} +c.c$ and $\rho = -i\trho e^{-i\omega_0 t}$, where 
the rotation wave approximation(RWA) 
is taken into consideration for the polarization. Applying the 
slowly varying approximation for time to Maxwell-Bloch Eqs. (\ref{Eeq}),
(\ref{Peq}), (\ref{Weq}), and (\ref{MB-Pz}) yields the
following equations:
%
%
\begin{eqnarray}
\dfrac{\partial}{\partial t}\tE
&=&
\dfrac{i}{2}
\left(
\dfrac{\partial^2}{\partial s^2}
+
\dfrac{n^2(s)}{n_0^2}
+
2i\dfrac{\Omega_D}{n_0}
\dfrac{\partial}{\partial s}
\right)\tE \nonumber \\
&-&\tilde{\beta}\tE
+
\eta(s) \trho + \tilde{F}_1(s,t),  \label{eq:SB-1}
\end{eqnarray}
\begin{eqnarray}
\dfrac{\partial}{\partial t}\trho
=
-(\tgamma_{\perp}+i\Delta_0)\trho
+
\tkappa W\tE + \tilde{F}_2(s,t), \label{eq:SB-2}
\end{eqnarray}
\begin{eqnarray}
\dfrac{\partial}{\partial t}W
=
-\tgamma_{//}(W-W_{\infty})
-2\tkappa(\tE\trho^* + c.c)
+ \tilde{F}_3(s,t), \label{eq:SB-3}
\end{eqnarray}
where space and time are made dimensionless by the scale transformation;
$(n_0\omega_0/c)s\rightarrow s$ and $\omega_0 t\rightarrow t$,
respectively, and
$\tilde{\beta}\equiv \beta c^2/(n_0^2\omega_0)$,
$\Omega_D\equiv R\Omega/c$, 
$\eta(s) \equiv (2\pi N_a\kappa\hbar/n_0^2)\Theta(s)$, 
$\tgamma_{\perp} \equiv \gamma_{\perp}/\omega_0$, 
$\tgamma_{//}\equiv \gamma_{//}/\omega_0$, 
$\tkappa \equiv \kappa/\omega_0$,
$\Delta_0 \equiv (\omega_0-\omega_t)/\omega_0$.
$\tilde{F}_i$ is noise with the strength of fluctuation 
$\tilde{\sigma}_i=c^2/(n_0^2\omega_0^2)\sigma_i$.

\begin{figure}
\begin{center}
\raisebox{0.0cm}{\includegraphics[width=6cm]{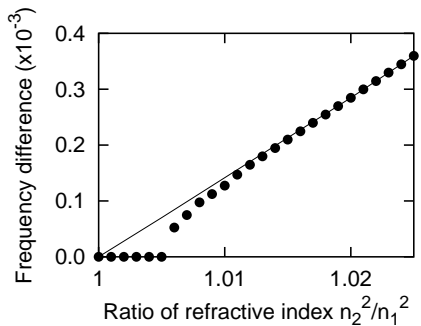}}
\end{center}
\vspace{-4mm}
\caption{\label{fig:neff09}
Frequency difference vs. 
ratio of refractive index $n_2^2/n_1^2$, where $n_1$ and $n_2$
 are respectively the refractive index of passive and active regions 
in a non-rotating ring laser ($\Omega=0$). 
Frequency difference is made dimensionless as
 $\Delta\omega_0/\omega_0$.
Solid line denotes frequency difference between eigenmodes
calculated following Eq. (\ref{f0}) in the ring cavity (without a gain 
medium), while closed circles are results 
obtained by numerical simulations of
 Eqs. (\ref{eq:SB-1})-(\ref{eq:SB-3}). 
Numerical results show that frequency-locking phenomenon occurs and frequency
 difference becomes zero when $n_2^2/n_1^2$ is less than $\sim 1.006$.
}
\end{figure}

In our simulations, we chose a ring cavity with (dimensionless) 
radius $R=10$ composed of a passive region and an active
region where there is a gain medium.
Then we suppose $n_1$ and $n_2$ to be the refractive
index of the passive and active regions, respectively. 
Length $l$ of the active region is set as
$l=(5\sqrt{2}\times 10^{-2})L$, where $L$ is the total cavity length $2\pi R$.
Then we set the values of most of the system's parameters as follows:
$\tkappa=0.5$, 
$W_{\infty} = 2\times 10^{-3}$,
$\tgamma_{\perp}=5\times 10^{-2}$,
$\tgamma_{//}=10^{-3}$,
$\tilde{\sigma}_j=10^{-7}$ $(j=1,2,3)$,
$\Delta_0=0$, 
and $\tilde{\beta}(s)=\alpha_L\Theta(s)$, where $\Theta(s)$ is a step
function that is 1 inside the active region and
zero outside it, and $\alpha_L=10^{-3}$. 
Here, note that the conservative backscatters 
are introduced by the variations of the refractive index in this cavity. 
Since the strength of backscattering $\gamma$ is related to the variation as $\sqrt{\gamma}\propto
(n_2^2/n_1^2-1)$,
 the resonant frequencies of the eigenmodes of a non-rotating cavity 
are split into two, following Eq. (\ref{f0}). 
The solid line in Fig. \ref{fig:neff09} 
shows the dependence of (dimensionless) 
frequency splitting $\Delta\omega_0/\omega_0$ on the ratio 
of the refractive index $n_2^2/n_1^2$.

For various values of $n_2^2/n_1^2$, 
we numerically integrate Eqs. (\ref{eq:SB-1})-(\ref{eq:SB-3}) 
to obtain two mode laser operation in the non-rotating ring laser 
($\Omega_D=0$). 
The detailed method of the numerical simulations is shown in Appendix
\ref{Appendix}.
 
In Fig. \ref{fig:neff09}, we show the $n_2^2/n_1^2$ dependence of 
the frequency difference between the two lasing modes by closed circles.  
As seen in this figure, it turns out that there is a zone corresponding
to the absence of the frequency difference for $n_2^2/n_1^2 \le 1.006$.
This is because of frequency-locking between two modes.
The regime ($n_2^2/n_1^2 \le 1.006$) 
corresponds to the case when strength $\gamma$
of conservative backscattering is smaller than critical value
$\gamma_c$. 
On the other hand, the frequency difference can be numerically observed  
without the frequency-locking phenomenon for $n_2^2/n_1^2>1.006$,  
and its value approaches the solid line obtained from 
Eq (\ref{f0}).
Below, we refer to the two regimes, $0<n_2^2/n_1^2 \le 1.006$ and 
$n_2^2/n_1^2>1.006$, as locking and unlocking regimes, respectively. 

\begin{figure}
\begin{center}
  \begin{tabular}{c}
\raisebox{0.0cm}{\includegraphics[width=6cm]{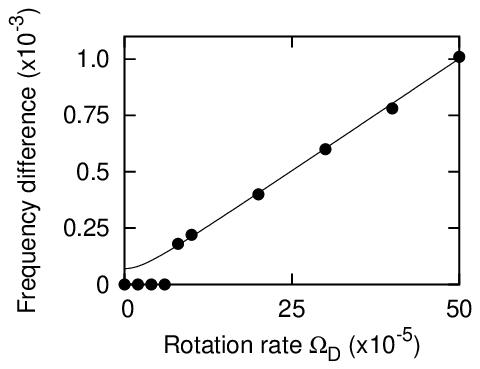}}
   \\ 
\hspace{1.2cm}(a)\\
\raisebox{0.0cm}{\includegraphics[width=6cm]{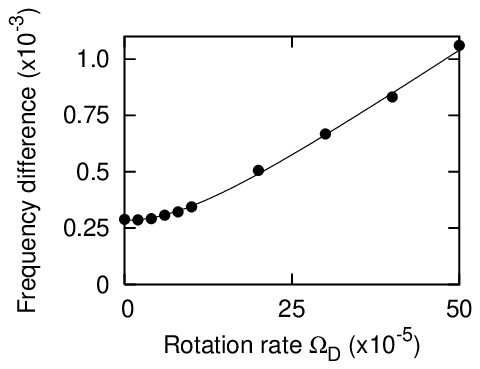}}  \\
\hspace{1.2cm}(b)
  \end{tabular}
\end{center}
\vspace{-4mm}
\caption{\label{fig:neff10}
(Dimensionless) rotation rate $\Omega_D$ vs. (Dimensionless) frequency
 difference $\Delta\omega_{\Omega}/\omega_0$ in rotating ring lasers
 with locking regime ($n_2^2/n_1^2=1.005$) (a) and unlocking
 regime ($n_2^2/n_1^2=1.02$) (b). 
Solid lines denote frequency difference calculated by
 Eq. (\ref{freqd}) in each case, while closed circles denote 
results obtained from numerical simulations 
of Eqs (\ref{eq:SB-1})-(\ref{eq:SB-3}).   
}
\end{figure}

Typical examples of the  rotation rate dependence of the frequency
difference are shown in Figs. \ref{fig:neff10}(a) and (b) for the
locking ($n_2^2/n_1^2=1.005$) and unlocking regimes
($n_2^2/n_1^2=1.02$), respectively.
The frequency difference 
calculated by Eq. (\ref{freqd}) in each case is represented by the solid
line. 
For the locking regime, when the ring laser is rotated but the rotation
rate is small, the frequency-locking phenomenon is still caused and 
therefore, the frequency difference between the two lasing modes 
cannot be observed [Fig. \ref{fig:neff10} (a)].
When rotation rate $\Omega_D$ is increased, however, the frequency
difference start to increase and approaches the values calculated by
Eq. (\ref{freqd}). 
The existence of the region corresponding to the absence of the
frequency difference might be qualitatively similar with so-called deadband \cite{Crow}. 
%
On the other hand, for the unlocking regime [Fig. \ref{fig:neff10} (b)], 
as mentioned above, the frequency-locking effect
is not caused even at zero rotation rate.  
As the rotation rate is increased, 
the frequency difference following Eq. (\ref{freqd}) increase without deadband, 
as shown in Fig. \ref{fig:neff10} (b). 
These numerical results confirm the theoretical
prediction that frequency difference can be observed without the
lock-in phenomenon in a ring laser with conservative backscattering. 

\section{Experiment \label{sec_exp}}
\begin{figure}
\begin{center}
\rotatebox{-90}
{
\raisebox{0.0cm}{\includegraphics[width=5cm]{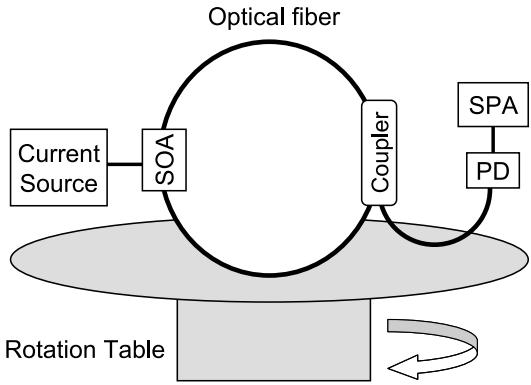}}
}
\end{center}
\vspace{-4mm}
\caption{\label{fig_setup}
Experimental setup of semiconductor fiber ring laser gyroscopes (S-FOG):
SOA, a semiconductor optical amplifier (its temperature is kept at 25
 degree by a thermo-controller); Coupler, 95:5 coupler ( 5 $\%$ 
of energy is extracted only from the CW-output signal inside ring
cavity.); PD, photo detector; SPA, rf spectrum analyzer.}
\end{figure}

Lastly, let us examine 
if the theoretical results can actually be obtained in a real
experiment. For this purpose, we used
a semiconductor fiber optic ring laser gyroscope (S-FOG)
\cite{Inagaki,Harayama}.  
The experimental setup is illustrated in Fig. \ref{fig_setup}. 
The ring laser part of S-FOG consists of a semiconductor optical
amplifier (SOA) as a gain medium and 
polarization-maintaining fibers to form a ring cavity.  
Conservative backscattering is mainly caused at the connection point
between the SOA and the optical
fibers in S-FOG, because there is an air region at the connection point, which leads to 
substantial changes of the refractive indices at the points \cite{complement}. 
From the return loss measurements, the value of $\gamma$ was approximately
estimated as $10^{-5}< \gamma < 10^{-4}$. 
%
%
The extraction of the laser output signal from the ring laser part is
achieved by a 95:5 coupler, where 
only 5$\%$ energy of the CW-output signal is extracted.
It is detected by photo detector
PD, as shown in Fig. \ref{fig_setup}.
SPA denotes a rf
spectrum analyzer that is used to observe the beat frequency signals 
of the output signal \cite{method}.  The entire system including power supplies
and measurement instruments is mounted on a rotation table.

\subsection{Estimation of $\gamma_c$}
To obtain the critical backscattering strength
$\gamma_c$ for the S-FOG, 
it is useful to estimate the parameters (dispersive coupling
$\gamma_{12}$ and linear net gain $g$) that
determine the $\gamma_{c}$- value (Eq. (\ref{eq:condition2})). 

Dispersive coupling $\gamma_{12}$ can be calculated on the basis of the definitions
shown in Table \ref{tab:Table1} as 
$\gamma_{12}\approx \Gamma/(2n_0kl)|\sin n_0kl\sin \theta|<
\Gamma/(2n_0kl)$, 
where $\Gamma$ is the total loss (typically, $10^{5}<\Gamma<10^{6}s^{-1}$) 
of the ring cavity.  
$k_0$ and $l$ are respectively the wavenumber of the light and 
SOA length ($l\approx 1.5$ mm), 
and $\theta$ is the parameter related to the
position of the SOA.  
For the calculations, we used the wavefunctions (\ref{pur_wf}) 
because $\gamma \ll 1$. 
Using the S-FOG's parameters (lasing wavelength
$\lambda\approx 1563.1$ nm, $n_0=1.445$, and $\Gamma\approx 10^{6}s^{-1}$)
with the above equation
yields dispersive coupling $\gamma_{12}<10^{4}s^{-1}$. 

Linear net gain $g$ of the SOA can be rewritten 
as $g=g_0(\alpha-1)$, where 
$\alpha$ is the normalized pumping power defined by
$\alpha=(W_{\infty}-W_{th})/W_{th}$ ($W_{th}$ is the pumping threshold), 
and coefficient $g_0$ 
is estimated as $g_0\approx 1.3\times 10^{7}(s^{-1})$ using the differential gain coefficient of the SOA.  
When we set normalized pumping strength $\alpha$ as $\alpha=1.06$ in
the ring laser of fiber length $L=4.12 m$, linear gain coefficient $g$ is estimated as 
$g\approx 6.6\times 10^5(s^{-1})$. 
Using this value and values $\gamma_{12}=10^{4}s^{-1}$, $K=5$ with
Eq. (\ref{eq:condition2}) yields critical backscattering strength
$\gamma_{c} \approx 7.2\times 10^{-6}$, which is smaller than 
the measured value of $\gamma(=10^{-5}-10^{-4})$. 
From these estimations, we expect that the S-FOG satisfies the condition
(\ref{eq:condition2}) and generates beat frequency signals
without the lock-in phenomenon. 
\subsection{Results and discussions}
\begin{figure}
\begin{center}
\raisebox{0.0cm}{\includegraphics[width=9cm]{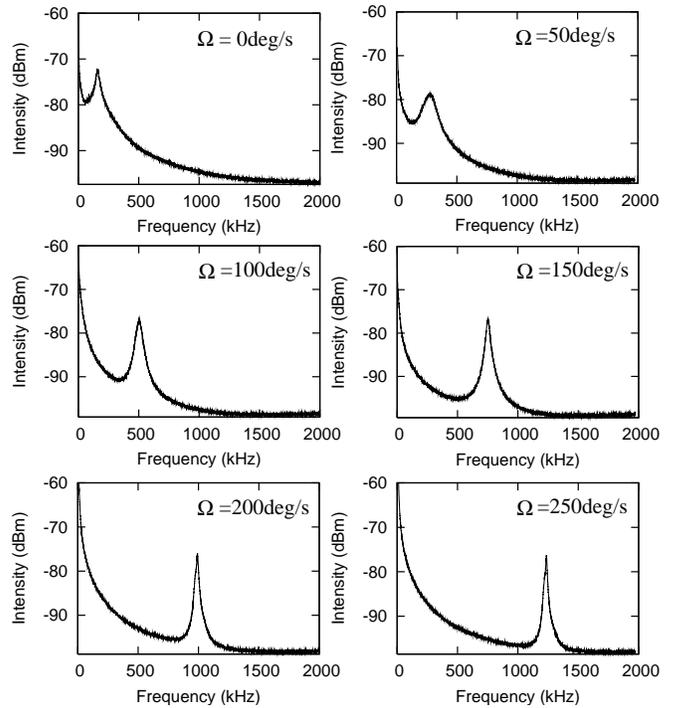}}
\end{center}
\vspace{-4mm}
\caption{\label{fig_rfspectrum}
rf spectrum at each rotation rate $\Omega$ (deg/s). Sharp peaks,
 representing beat frequencies, can be observed in $1/f$-like noise.
}
\end{figure}

\begin{figure}
\begin{center}
\raisebox{0.0cm}{\includegraphics[width=9cm]{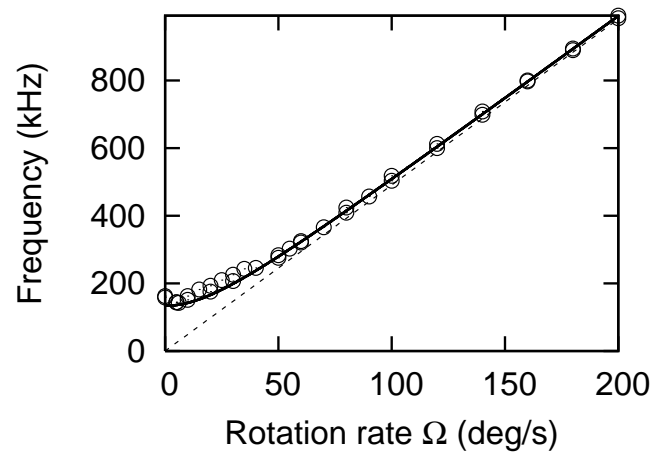}}
\end{center}
\vspace{-4mm}
\caption{\label{fig_frequency_response}
Beat frequency vs. rotation rate. Open circles denote 
beat frequency obtained from experiments, while dashed line
 denotes an asymptotic line.
Eq. (\ref{freqd}) is represented by solid line. 
}
\end{figure}

%
Figure. \ref{fig_rfspectrum} shows the measured rf spectra obtained from
the PD output signal at normalized pumping power $\alpha=1.06$,
where the spectrum is time-averaged to obtain clear beat signals.  
As predicted by the above estimation, a peak corresponding
to the beat frequency can be seen in the rf-spectrum even at zero rotation rate 
($\Omega=0 deg/s$).
Then the peak frequency is changed as the rotation rate is increased. 
Fig. \ref{fig_frequency_response} shows 
the rotation rate dependence of the peak frequencies.
The experimental result 
demonstrates highly linear characteristics only in cases of high rotation rate, while
there is substantial deviation from
an asymptotic line given by the dashed line for slow rotation rates.

\begin{figure}
\begin{center}
\raisebox{0.0cm}{\includegraphics[width=9cm]{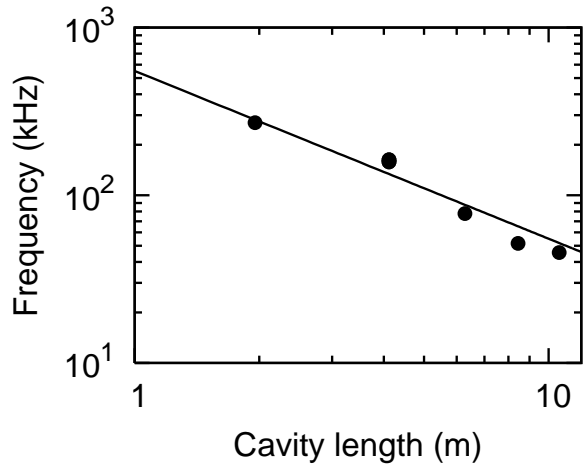}}
\end{center}
\vspace{-4mm}
\caption{\label{fig_f0L}
Beat frequency observed at rotation rate
 $\Omega=0$ (deg/s) vs. cavity length. Closed circles denote
 beat frequency at zero rotation rate. Eq. (\ref{f0}) is represented
 by a solid line.
}
\end{figure}

In addition, we confirmed that
 the similar rotation rate dependence with that shown in 
Fig. \ref{fig_frequency_response}
can be observed even when changing the length of an optical fiber 
of the ring cavity part. 
According to Eq. (\ref{f0}), the beat frequency observed at zero rotation rate 
($\Omega=0$) can be controlled by changing the fiber length.  
In Fig. \ref{fig_f0L}, we show the dependence of the beat frequency 
on the fiber length. We find that the dependence can be explained by using
 Eq. (\ref{f0}) with the assumed backscattering strength
 $\gamma=7.0\times 10^{-5}$, which is given by the solid line in Fig. \ref{fig_f0L}.  
Assumed $\gamma$- value is consistent with 
the order of the value $10^{-5}-10^{-4}$ obtained from the return loss measurement.
Moreover, to explain the rotation rate dependence of the beat
frequency shown in Fig. \ref{fig_frequency_response} by using the assumed
 $\gamma$ -value, we superimpose Eq. (\ref{freqd}) with $\gamma=7.0\times
10^{-5}$ in the solid curve of Fig. \ref{fig_frequency_response}.   
It also turns out that the tendency of the experimental results is
well reproduced by Eq. (\ref{freqd}).
Consequently, we conclude that we observed the avoidance of the lock-in
 phenomenon for $\gamma>\gamma_c$ and the manifestation of  
 the Sagnac effect (\ref{freqd}) incorporating the effect of
 backscattering in the experiment of the S-FOG.

\section{Summary \label{sec_sum}}
In summary, we studied the effect of backscattering on rotating ring
lasers by using the Maxwell-Bloch Eqs. 
(\ref{Eeq}), (\ref{Peq}), (\ref{Weq}), and (\ref{MB-Pz}) 
and showed that the occurrence of the lock-in phenomenon can be
characterized by the strength of the conservative backscattering 
that originates in the bumps of the refractive index. 
Particularly, we revealed the existence of critical backscattering
strength $\gamma_c$, above which the Sagnac frequency shifts described
by Eq. (\ref{freqd}) can be observed without the lock-in phenomenon.
We also showed that 
the experimental results corresponding to the theoretical one is 
actually obtained in a S-FOG. 
We believe that the property that lock-in phenomenon does not occur 
is not unique for S-FOGs but universal for RLGs using solidstate and
semiconductor as gain medium. 

\begin{acknowledgments}
We thank Professor Ezekiel, Mr. Ohno, and Professor Hotate 
for fruitful discussion and valuable comments on S-FOGs.
We also thank T. Miyasaka for numerical simulations and
S. Shinohara for discussions.  
This work was supported by 
the National Institute of information and Communication Technology
of Japan. 
\end{acknowledgments}

\appendix 
\section{Numerical simulation method \label{Appendix}}
%
%
Let us explain the numerical simulation method of 
Eqs. (\ref{eq:SB-1})-(\ref{eq:SB-3}).
The approximated Maxwell equation (\ref{eq:SB-1}) has identical form 
to the Schr\"odinger equation with absorption $\tilde{\beta}$, 
polarization $\trho$, and noise $\tilde{F}_1$; 
it allows us to use the techniques of
the symplectic-integrator method \cite{TakahashiIkeda} to carry out
numerical simulations of dynamics. 

We divide Eq.~(\ref{eq:SB-1}) into  
 wave propagation, polarization, and noise parts.
The time evolution operator of the wave propagation part is given 
as $U_{\tau} = e^{-i\hat{H}\tau}$, where $\hat{H}$ consists of the
kinetic energy part 
$\hat{T}=
-1/2(\partial^2/\partial s^2 +2i\Omega_D/n_0\partial/\partial s)$ 
and the potential part including linear absorption term 
$\hat{V}=-n^2/(2n^2_{0})-i\tilde{\beta}$ as 
$\hat{H}=\hat{T}+\hat{V}$.
If time $\tau$ is very small, the time-evolution operator  
can be split based on the symplectic-integrator method as follows:
 \begin{equation}
U_{\tau}=e^{-i\hat{V}\tau/2}e^{-i\hat{T}\tau}e^{-i\hat{V}\tau/2}+O(\tau^3).
\end{equation}
The next time-step value of electronic field
$\tE(t+\tau,s)$ is calculated by operating the split $U_{\tau}$ on the
previous time-step value $\tE(t,s)$ and including the contribution
of the polarization term and a noise term 
up to the first order correction of $\tau$:
\begin{eqnarray} 
\tE(t+\tau,s) \approx U_{\tau} \tE(t,s) + 
\eta(s) \trho(t,s)\tau + df_1(s,t), 
\label{SIMU-E}
\end{eqnarray}
where $df_1=\tilde{F}_1\tau$. 

Equation (\ref{SIMU-E}) can be calculated by the following procedures;
First, potential part $e^{-i\hat{V}\tau/2}$ is
operated on $\tE(t,s)$, ($\tE_1(t,s)\equiv e^{-i\hat{V}\tau}\tE(t,s)$). 
Then the Fourier translation of $\tE_1(t,s)$
is calculated as $\hat{E_1}(t,k)=\oint \tE_1(t,s)e^{-iks}ds$.
Operating $\hat{E_1(t,k)}$ to $e^{-i\hat{T}\tau}$ yields the following result:
\begin{eqnarray} 
\tE_2(t,s)
= \frac{1}{2\pi}\int e^
{-i/2
\left(
k^2+2\Omega_D/n_0 k
\right)\tau}
\hat{E_1}(t,k)e^{iks}dk.
\end{eqnarray}
Then $e^{-i\hat{V}\tau/2}\tE_2(t,s)(\equiv \tE_3(t,s))$ is obtained.
Finally, the next time step $\tE(t+\tau,s)$ 
is obtained by adding polarization
term $\eta(s)\trho(t,s)\tau$ and noise term $df_1(t,s)$ 
to $\tE_3(t,s)$.
$df_1$ is generated 
by converting uniform random numbers generated by the M-sequence technique
to normal distribution random numbers by using Box-Muller's formula. 

For the Bloch equations, the next time-step value of polarization
$\trho(t+\tau,s)$ and population inversion $W(t+\tau,s)$ 
are calculated by the Euler method:
\begin{eqnarray} 
\trho(t+\tau,s)=\trho(t,s)  
+\dfrac{d}{dt}\trho(t,s)\tau + df_2(t,s),
\label{Disrho}
\end{eqnarray}
\begin{eqnarray} 
W(t+\tau,s)=W(t,s)
+\dfrac{d}{dt}W(t,s)\tau + df_3(t,s),
\label{DisW}
\end{eqnarray}
where $df_i(t,s)=\tilde{F}_i(t,s)\tau$ $(i=2,3)$ are also generated by the 
above method.

%


\begin{thebibliography}{99}
\bibitem{Neelen_ex}
In the case of $|\gamma_{12}|\gg g$, 
$\Delta\omega_c$ is similar with a criterion used in
	Ref. \cite{Neelen}, above which an intensity
oscillation of CW- and CCW- rotating waves in antiphase occurs due to
	the two simultaneously oscillating standing-wave modes.
\bibitem{complement}
Except for the points, 
the other connection points in our ring cavity are between optical fibers.
However, angled physical contact (APC) connectors are used at the
points, and the return loss is less than $10^{-6}$, which can be omitted
compared to the return loss at the connection points between SOA 
and the optical fibers.
\bibitem{method}
This is because beat frequency $\Delta\omega_{\Omega}$ is transformed to
the amplitude oscillation due to the modulation effects of
Eq. (\ref{mode-eq1}). 
Thus, an extra directional coupler is not needed to combine
the extracted CW- and CCW- output signals in order to produce 
the beat signals. 
It has been experimentally confirmed that the amplitude oscillation frequency agrees very well with
beat frequency measured by the conventional method with an extra coupler\cite{Harayama}.
\end{thebibliography}
\end{document}